\documentclass[a4paper,fleqn]{cas-sc}

\usepackage[authoryear,longnamesfirst]{natbib}
\graphicspath{ {Figures/} }
\usepackage{enumitem}

\begin{document}
\let\WriteBookmarks\relax
\def\floatpagepagefraction{1}
\def\textpagefraction{.001}
\shorttitle{}
\shortauthors{K Sharma et~al.}

\title [mode = title]{Patterns of retractions from 1981-2020 : Does a fraud lead to another fraud?}                      

%

\author[1]{Kiran Sharma} \corref{cor1}
\ead{kiransharma1187@gmail.com}
\credit{Conceived and designed the analysis; Collected the data; Contributed data or analysis tools; Performed the analysis; Writing - original draft}
\address[1]{Chemical \& Biological Engineering, Northwestern University, Evanston, Illinois-60208, USA}

\begin{abstract}
Misconduct accounts for the majority of retracted scientific publications and this database reveals the disturbing trend in science~\citep{fang2012misconduct, brainard2018massive}.
The objective of the study is to find the association among the authors' collaboration, the number of retracted papers, the number of retracted citations,  journal impact factor, and research areas.
We present a detailed analysis of 12231 research papers indexed by Web of Science (WoS) as retracted publications from 1981-2020. The study demonstrates the collaboration patterns of retracted publications where 61.5\% of authors have only one and 24.6\% have two retracted papers; however, 2\% of authors have more than 10 retracted papers. To study the impact of citing retracted papers, we investigated the retracted papers with citations. The study reveals that 55.2\% of retracted papers have been cited at least once, where 25.4\% of papers are such papers where at least one citation turned out to be a retraction. This shows the impact of scientific misconduct or fraud on new research.
The number of retractions is independent of the journal impact factor and as compared to high impact papers, low impact papers are attracting more citations. We also investigate the citations received by retracted papers published in higher as well as lower impact factor journals. 1/4th of the papers are retracted citations that cited the retracted papers; however, there is no significant relationship exists between the higher impact or lower impact journals with retractions or citations.
Finally, how the average team size and average retracted citations vary among different research areas are studied. The study provides an insight that how a fraud leads to another fraud in the scientific world. Also, the rising trend of citations of retracted papers is a serious concern.

\end{abstract}


\begin{highlights}
\item The objective of the study is to investigate the increasing collaborations of retracted papers and citations of such papers. We have analyzed 12231 retracted papers indexed by Web of Science from 1981-2020. 

\item We have examined the temporal evolution of collaboration patterns along with the authors' contribution.  We have observed that teams smaller in size have more retractions.

\item We have demonstrated the impact of retracted papers on the papers citing the retracted papers.  We have analyzed that 1/4th of the papers are retracted citations which referred to the retracted papers.

\item We have observed the number of retractions and citations are independent of journal impact factor; however, low impact retracted papers keep on attracting more citations.

\end{highlights}

\begin{keywords}
Retracted publications \sep Scientific misconduct  \sep Bibliometric analysis \sep Collaboration patterns \sep 
\end{keywords}

\maketitle
\section{Introduction}
\label{sec:Intro}

In 2000,  Jan Hendrik Sch\"{o}n became famous for his work on coax materials into superconductors, and he published eight papers in Science and Nature.  Julia Hsu and Lynn Loo accidentally stumbled across duplicated data used in one of the Sch\"{o}n's paper while verifying the experimental progress of their work. That raised an alarm bell to bring the attention of the scientific community towards the Sch\"{o}n's work and they found evidence of scientific misconduct in at least 16 of them.  It raised questions on publication policies of these high impact journals too.
Scandals like Hwang Woo-Suk's fake stem-cell lines~\citep{saunders2008research} or Jan Hendrik Sch\"{o}n's duplicated graphs~\citep{service2003more} showed how easily scientist can publish fabricated data and put human health at risk. along with wasting millions of dollars of government research money.
These scandals raised questions on the role and responsibility of coauthors and reviewers of scientific articles. Similarly,  a Japanese bone-health researcher, named Yoshihiro Sato,  fabricated data, plagiarized work,  and forged authorships in more than 60 studies from 1996 to 2013~\citep{kupferschmidt2018researcher, else2019universities}.
There are multiple reasons behind the retraction of the papers including duplicate publication, falsification of the data, fabrication of the data, plagiarism, ethical or legal misconduct, etc. \citep{steen2013has, bohannon2016s}. Some times back Chinese journal found 31\% of submissions plagiarized~\citep{zhang2010chinese, mallapaty2020china}. \cite{grieneisen2012comprehensive} have discussed the reasons and proportion of retractions as research misconduct (20\%), questionable data or interpretation (42\%), and publishing misconduct (47\%). 
A published research work completely depends on the trust that a reader built with a writer where a reader expects honesty and fairness in the results reported by writer~\citep{rennie1997authorship}.
While publishing any of the research work, authors must be prepared to accept the public responsibility that goes with it, i.e.,	 authors are certifying the integrity of their work~\citep{shapiro1994contributions}. 

With the rise in scientific collaborations, number of retractions also increased from last two decades. Although scientific collaborations increases the research productivity by using individual's knowledge and research skill set; however, a wrong collaboration can create a distrust among peers and spoil the reputation~\citep{abramo2011relationship, mongeon2016costly}. \cite{bennett2003unethical} have shown in their study that how unethical practices in authorship of scientific papers have increased over the past few decades.
\cite{khan1999controlled} have explained in their study the rising trend in multiple authorship and this trend is sharp in medical papers~\citep{onwude1993multiple, king2018analysis}. Most of the fraud is happening in the medical field and many studies have highlighted the scientific misconduct in medicine~\citep{steen2011retractions, cassao2018retracted}.
There are large number of authors who keeps on publishing fraud research~\citep{fanelli2009many, kuroki2018repeating}. This raise a serious concern on authors that are they deliberately committing research fraud?~\citep{steen2011retractions}.
\cite{halevi2016post} have explored the post retraction citations to retracted papers and found that the huge majority of the citations are positive, and the citing papers usually fail to mention that the cited article was retracted. 
Earlier studies highlight that journal's retraction rate and its retractions for fraud are positively associated with the journal's impact factor~\citep{fang2011retracted, fang2012misconduct}. \cite{trikalinos2008falsified} stated in their study that falsified papers in high-impact journals were slow to retract. Nowadays, high impact journals have revised their retraction policies\citep{atlas2004retraction, resnik2015retraction} which lower down the number of retractions.

Over the years, a number of studies has been conducted to explain the characteristics of scientific misconduct \citep{he2013retraction, bar2018temporal}, the rise of
retractions and its causes \citep{cokol2008retraction, steen2011retractions}, effect of collaborations \citep{franceschet2010effect, asubiaro2019collaboration, zhang2020collaboration, sharma2020growth}, behavior of author \citep{martinson2005scientists}, impact on authors' career~\citep{mongeon2016costly}, the ongoing citations of retracted papers \citep{neale2010analysis}, highly cited retracted papers~\citep{da2017highly}, etc.  Our study provides a systematic view on the growth of the team sizes in retracted papers, a relationship between the authors' collaboration and retractions, an association between retractions and retracted citations, a relationship between journal impact factor and retracted citations, and disciplines with higher retractions and team size.
The study is organized as follows: Section~\ref{sec:data} describes the data. Results are explained in Section~\ref{sec:result}. Finally, the concluding remarks are presented in Section~\ref{sec:conclusion}.
\section{Data description}
\label{sec:data}

We have downloaded the data from the Web of Science (WoS) managed by Clarivate Analytics (\url{https://apps.webofknowledge.com/}). We searched for papers which have \textit{Document Type} in WoS as \textit{Retracted Publication} or \textit{Retraction} from 1981-2020. The given condition filtered 12231 records. The metadata contains a paper unique ID,  author's name, affiliation,  journal name,  document type, research area, number of citations, etc.
The metadata contains information of 27822 authors, 3127 journals, and 146 research areas.
Further, to retrieve \textit{Document Type} of all citations respective to papers, first we check for papers with at least one citation and filtered 6754 papers. Further, we searched for every citation received by all 6754 papers in WoS and checked whether the cited paper is documented as either \textit{Retracted Publication} or \textit{Retraction}. This way we have filtered 3110 papers which has at least one citation documented as a retracted publication.

\section{Results}
\label{sec:result}

\subsection{Retractions and authors' collaboration}
\begin{figure}[!h]
    \centering
    \includegraphics[width=0.95\linewidth]{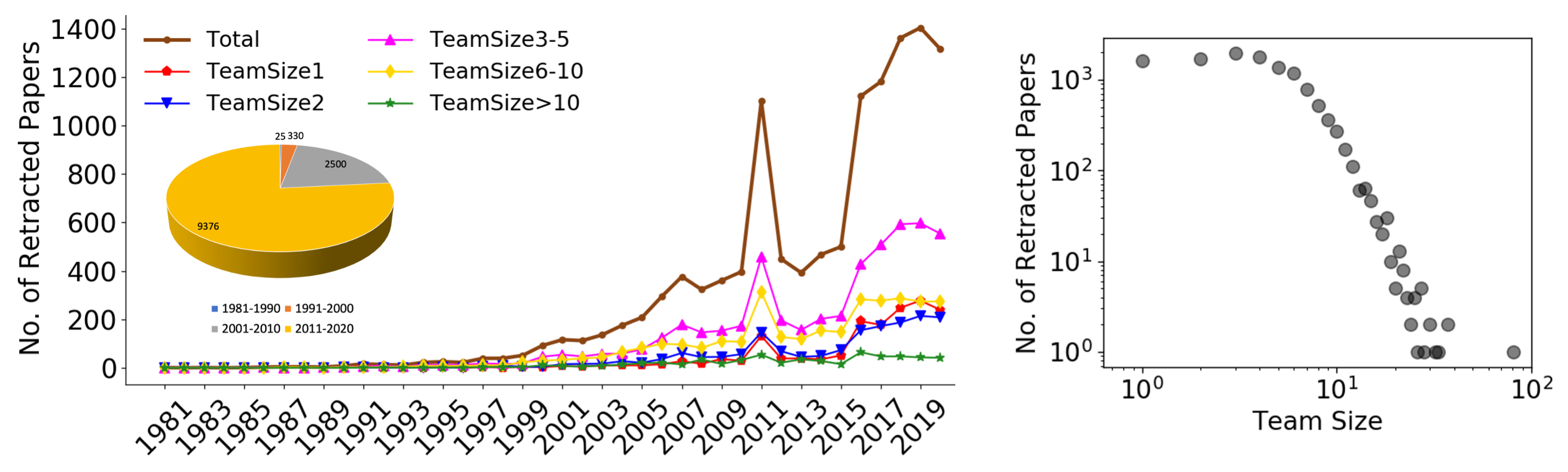}
        \llap{\parbox[b]{6.2in}{(a)\\\rule{0ex}{1.7in}}}
     \llap{\parbox[b]{2.4in}{(b)\\\rule{0ex}{1.7in}}}
\caption{(a) The number of papers retracted from 1981-2020. The trend line colored in brown represents the total number of retracted papers.Other trend lines show the number of authors (team size) in retracted papers. The teams of size 3-5 have a large number of retractions.  Also, the decade wise number of retractions is shown in inset. In 2001-2010, $20.4\%$ and in 2011-2020, $76.7\%$ of papers are retracted. (b) A number of retracted papers corresponding to varying team sizes. The minimum team size is a team of 2 authors and the maximum is a team of 81 authors. The highest number of papers were retraced from a team of size 1-5. }
\label{fig:Fig1}   
\end{figure}

The growth of the number of retracted papers as per the number of authors (team size) from 1981-2020 is shown in Fig.\ref{fig:Fig1}(a). The trend is shown for teams of size 1, 2, 3-5, 6-10, $>$10, and the total number of retractions. There is a sharp rise and fall in number of retractions from 2010-2012.  During 2010-2011 there is a rise in the number of retractions and then a fall is noticed immediately after in 2012.  Further, during 2015-16 again a sharp rise is happened in the number of retractions.
Decade-wise number of retractions; 1981-1990 ($0.2\%$), 1991-2000 ($2.7\%$) 2001-2010 ($20.4\%)$, and 2011-2020 ($76.7\%$) is shown in Fig.\ref{fig:Fig1}(a) inset. The papers written by more than 3 and less than 10 authors in collaborations have a large retraction count. Fig.\ref{fig:Fig1}(b) also shows that larger teams have a lesser number of retracted papers as compared to shorter teams. Single author papers have 13.3\% and two authors' papers have 13.9\% of retractions. Teams of size 3-5 have 42.3\% and teams of size 6-10 have 25.7\% of retracted papers. The percentage of retracted papers decreases with increase in team size while 69.5\% of retraction is from teams smaller than 5 authors. This highlights that a larger number of authors write individuals or tend to collaborate with a few others. \cite{steen2013has} has also studied the growth of scientific retractions and authors with multiple retractions; however, we demonstrated the authors' collaboration patterns in retracted papers.

\begin{figure}[!h]
    \centering
      \includegraphics[width=0.85\linewidth]{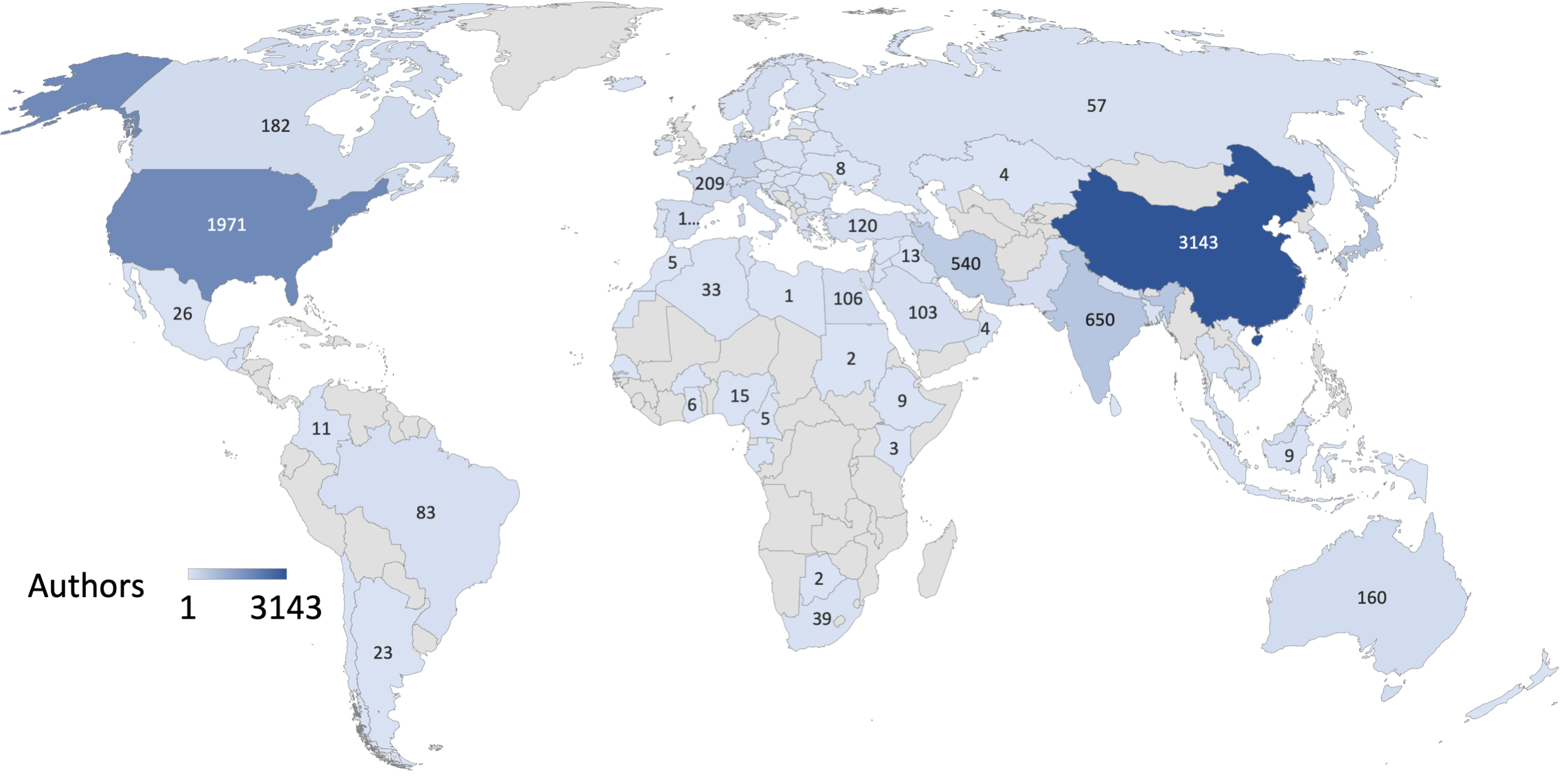}
      \llap{\parbox[b]{5.5in}{(a)\\\rule{0ex}{2.0in}}}
    \includegraphics[width=0.75\linewidth]{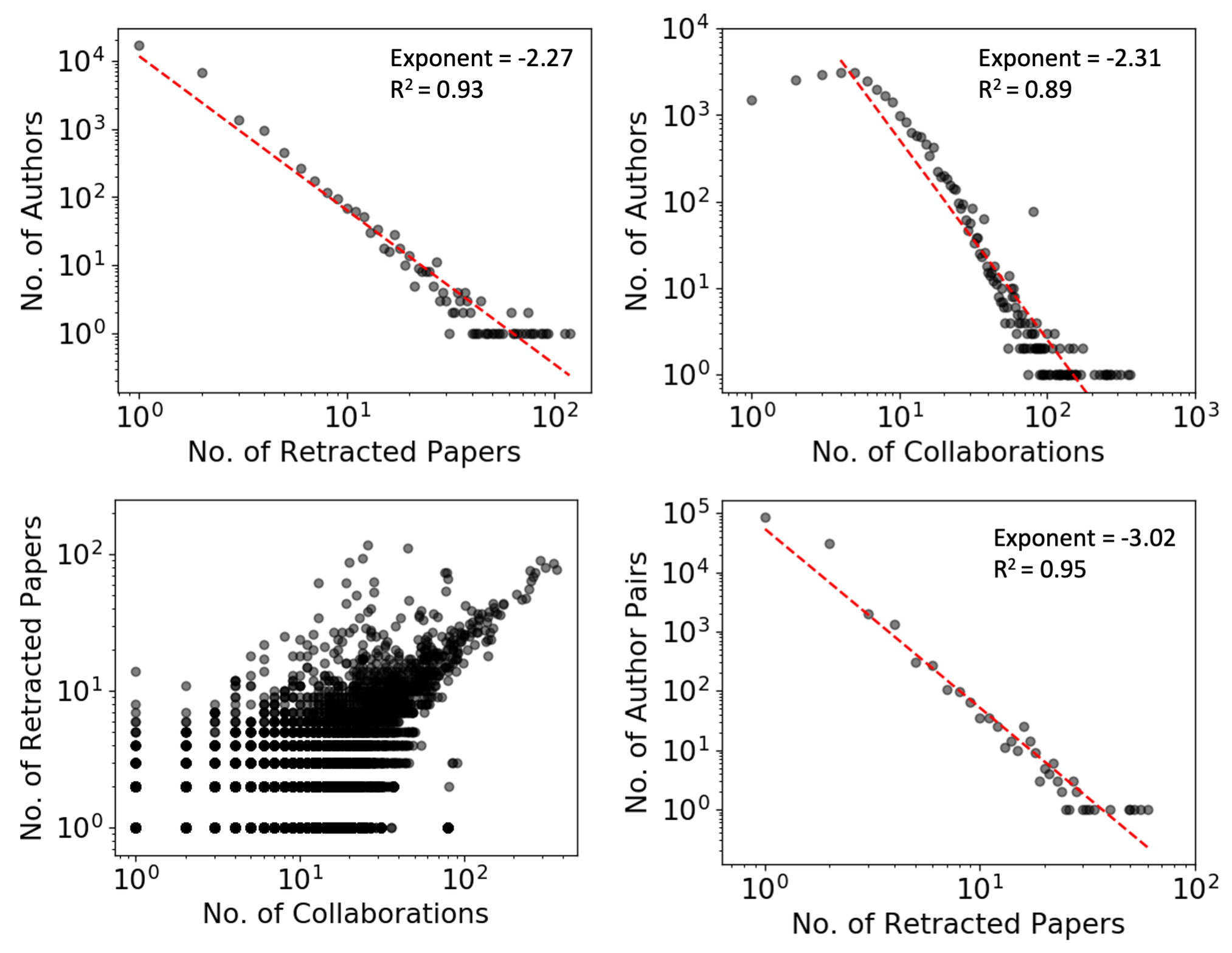}
    \llap{\parbox[b]{5.0in}{(b)\\\rule{0ex}{3.7in}}}
     \llap{\parbox[b]{2.5in}{(c)\\\rule{0ex}{3.7in}}}
     \llap{\parbox[b]{5.0in}{(d)\\\rule{0ex}{1.9in}}}
   \llap{\parbox[b]{2.6in}{(e)\\\rule{0ex}{1.9in}}}
\caption{(a) Geolocation of authors with a number of retractions. China has produced a large number of authors with retracted papers followed by the USA, India, Japan, and Iran with more than 500 authors.  The grey area in the map specifies no record. (b) The number of papers retracted by authors follows a power law,  $f(x) = x^{-k}$ with exponent -2.27 and $R^2$ = 0.93. (c) The number of collaborations shared by authors follows a power law with exponent -2.31 and $R^2$ = 0.89. (d) Authors with a number of collaborations and a number of retracted papers. (e) The number of papers retracted by pair of authors follows a power law with exponent -3.02 and $R^2$ = 0.95. A few author pairs have a higher retraction rate.}
\label{fig:Fig2}       
\end{figure}

Geolocation of authors with the number of retreated papers is shown in Fig.\ref{fig:Fig2}(a). Authors from China have higher retractions (25.7\%), followed by USA (16.1\%), India (5.3\%), Japan (5.2\%), Iran (4.4\%), Germany(3.2\%), England (2.8\%), South Korea (2.6\%), Italy (2.4\%) and so on. The co-authorship network has 27822 authors and 167645 collaborations among authors. The relationship among the number of authors and retracted papers follows a power law with exponent $-2.27$ as shown in Fig.\ref{fig:Fig2}(b). Out of 27822 authors, $61.5\%$ of authors have owned only one, and $24.6\%$ have owned two retracted papers; however, $65.4\%$ of retracted papers are owned by 27 individual's ($0.1\%$).
The share of collaborations among authors also follows a power law with exponent $-2.31$ as shown in Fig.\ref{fig:Fig2}(c). A large number of authors worked within a closed group. The authors who share large collaborations also have large number of retracted papers as shown in Fig.\ref{fig:Fig2}(d). The pair of authors with a number of retracted papers follows a power law with exponent $-3.02$ as shown in Fig.\ref{fig:Fig2}(e). 28.8\% of author pairs have more than one retracted paper. This analysis shows how the same set of authors and pairs of authors committing fraud again and again. Also, the relationship among authors and retractions shows a scale-free behavior with exponent varies between 2 to 3.

\subsection{Retracted citations}
\begin{figure}[!h]
    \centering
    \includegraphics[width=0.85\linewidth]{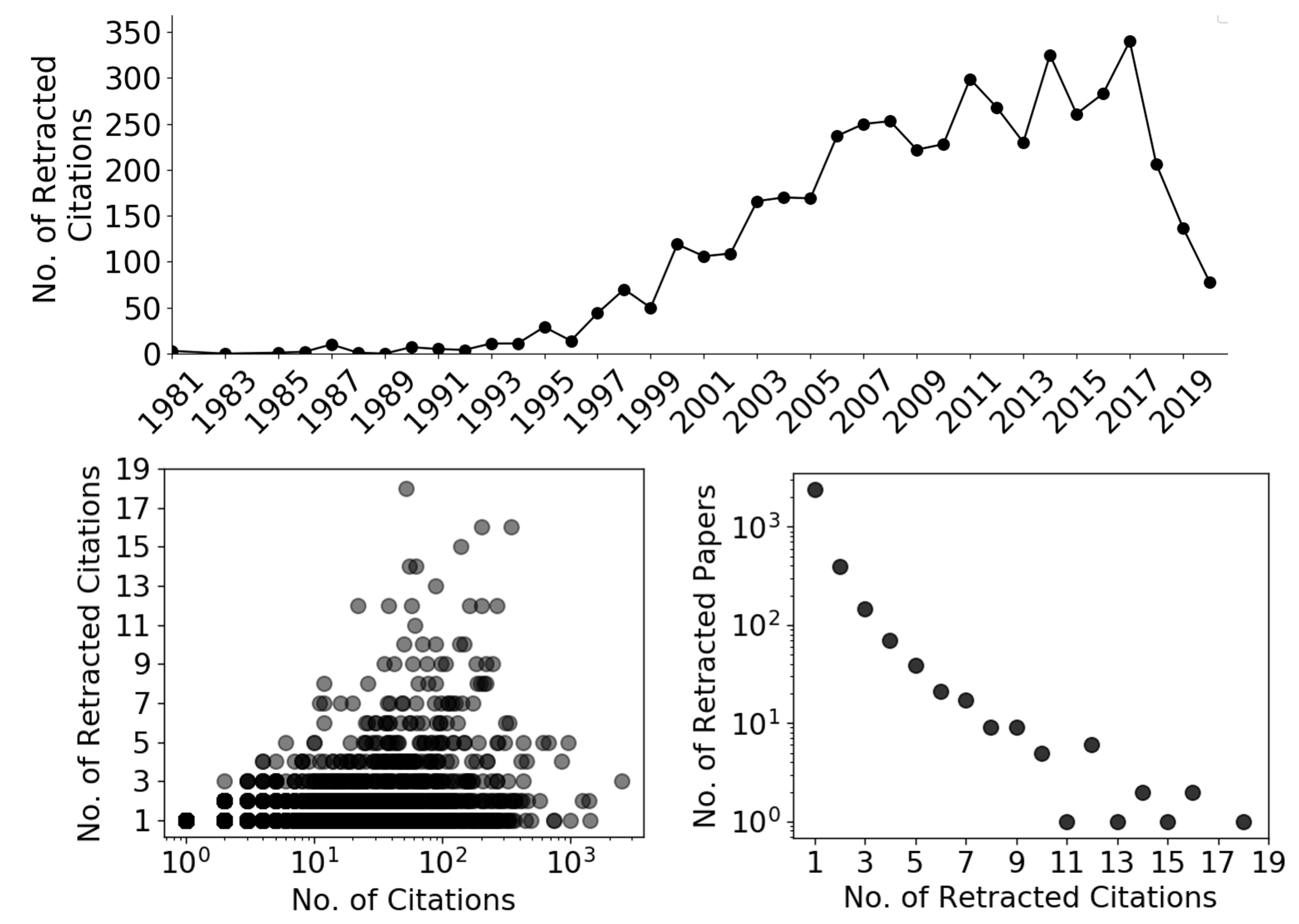}
       \llap{\parbox[b]{5.3in}{(a)\\\rule{0ex}{3.9in}}}
    \llap{\parbox[b]{5.5in}{(b)\\\rule{0ex}{2in}}}
     \llap{\parbox[b]{2.65in}{(c)\\\rule{0ex}{1.9in}}}
     
\caption{(a) The number of retracted citations from 1981-2020.  (b) The relationship between the number of citations received by retracted papers and the number of retracted citations. (c) The relationship between the number of retracted citations and the number of retractions. A few papers have a large number of retracted citations.}
\label{fig:Fig3}    
\end{figure}

\cite{da2017highly} have mentioned in their study why citing retracted papers can be a problem for academia. \cite{da2017some} also tried to explain the reason behind the retracted citations.
We studied the effect of citing wrong research by analyzing the citations of retracted papers.
55.2\% (6754) of retracted papers have received at least one citation. In contrast, 25.4\% (3110) papers are those whose citations have at least one retraction. Fig.\ref{fig:Fig3} shows the analysis of 3110 retracted papers where citations are retracted papers. The cumulative growth of the number of retracted citations from 1981-2020 is shown in Fig.\ref{fig:Fig3}(a). The relationship between the number of citations received by retracted papers and the number of retractions on those citations is plotted in Fig.\ref{fig:Fig3}(b). There are a few papers with a large number of retracted citations as shown in Fig.\ref{fig:Fig3}(c).  There is one such paper where 33.3\% of citing papers are retractions.  There could be two reasons behind this pattern: (i) People who cited the paper was unaware of the authenticity of the work, or the cited paper had been retracted later; (ii) The authors who cited the retracted papers are one of the authors from the retracted papers. However, the aim of the paper is not to look into the authors' profiles.

\subsection{Journal impact factor and research areas}

\begin{figure}[!h]
    \centering
    \includegraphics[width=0.85\linewidth]{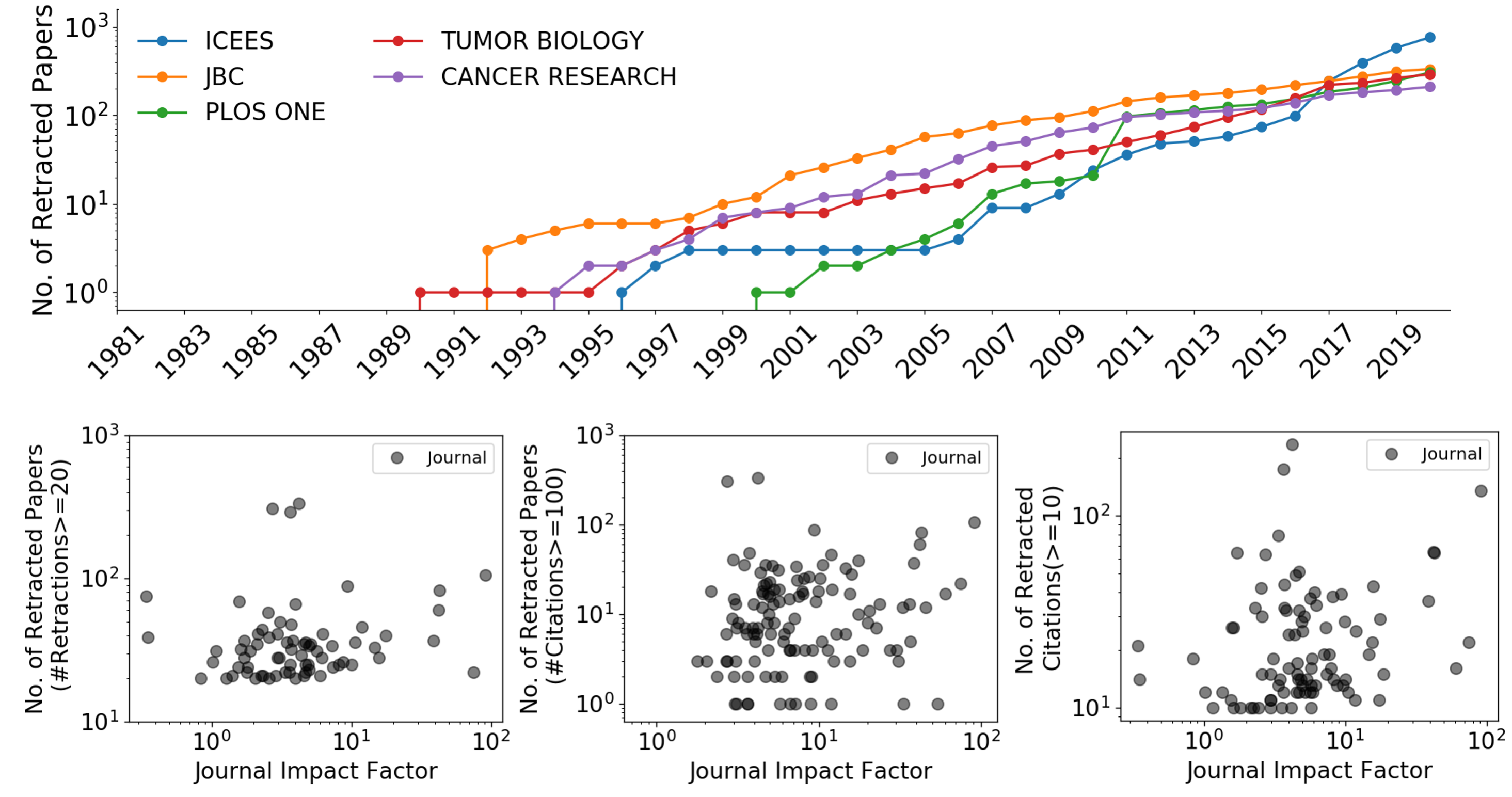}
     \llap{\parbox[b]{5.65in}{(a)\\\rule{0ex}{2.73in}}}
    \llap{\parbox[b]{5.6in}{(b)\\\rule{0ex}{1.33in}}}
     \llap{\parbox[b]{3.8in}{(c)\\\rule{0ex}{1.3in}}}
      \llap{\parbox[b]{2.0in}{(d)\\\rule{0ex}{1.25in}}}
\caption{(a) The cumulative growth of retractions in journals with more than 100 retracted papers. Relation of journal impact factor to retractions for (b) number of retracted papers $\geq 20$; (c) number of citations $\geq 100$. (d) Relationship between journal impact factor and the number of retracted citations. The number of retractions, citations, and retracted citations are independent of the journal impact factor. }
\label{fig:Fig4}    
\end{figure}

Fig.\ref{fig:Fig4}(a) shows the cumulative growth of the number of retracted papers in journals having more than 100 retractions since 1981. \textit{International Conference on Energy and Environmental Science 2011} (ICEES) has
a large number of retracted papers (6.2\%) with an average citation of 0.24, whereas \textit{Journal Biological Chemistry} (JBC) has 2.7\% of retracted papers with an average citation of 28.4. Details of the journals based on the number of retractions (>100) are listed in Table\ref{tabel:journal}. Fig.\ref{fig:Fig4}(b) shows the scattered plot of journals with more than 20 retractions. There is no significant relationship exists between the journal impact factor and the number of retractions. Similarly, the scattered plot of journals where the retracted papers have received more than 100 citations is shown in Fig.\ref{fig:Fig4}(c). The highly cited retracted papers have no significant relationship with the journal impact factor too.
Further, we investigated the citations received by retracted papers published in higher as well as lower impact factor journals. 1/4th of the papers turned out to be the retractions that have cited the retracted papers; however, there is no significant relationship exists between the higher or lower impact journals with the number of retractions or citations (see Fig.\ref{fig:Fig4}(d)). As we have seen in Fig.\ref{fig:Fig4} that the number of retractions is independent of the journal impact factor; however, all journals accept paper after a peer-reviewing. This raises a serious question on the work done by the scientific community before accepting the paper for publication.  How responsible are the journal and team associated with the journal for the fraud research?
\begin{table}[!h]
\caption{Journals with more than 100 retracted papers from 1981-2020.}

\begin{tabular}{|l|c|c|c|c|}
\hline
Journal                         & \multicolumn{1}{l|}{Impact Factor} & \multicolumn{1}{l|}{\#Retraction} & \multicolumn{1}{l|}{Avg. Citations} & \multicolumn{1}{l|}{Avg. Team Size} \\ \hline
ICEES                           & -                                  & 760                               & 0.24                                & 3.0                                   \\ \hline
JOURNAL OF BIOLOGICAL CHEMISTRY & 4.23                               & 334                               & 28.4                                & 6.5                                 \\ \hline
PLOS ONE                        & 2.74                               & 306                               & 4.8                                 & 4.7                                 \\ \hline
TUMOR BIOLOGY                   & 3.65                               & 291                               & 6.9                                 & 4.6                                 \\ \hline
CANCER RESEARCH                 & 91.3                               & 106                               & 46                                  & 6.6                                 \\ \hline
\end{tabular}
\label{tabel:journal}
\end{table}

\begin{figure}[!h]
    \centering
    \includegraphics[width=0.85\linewidth]{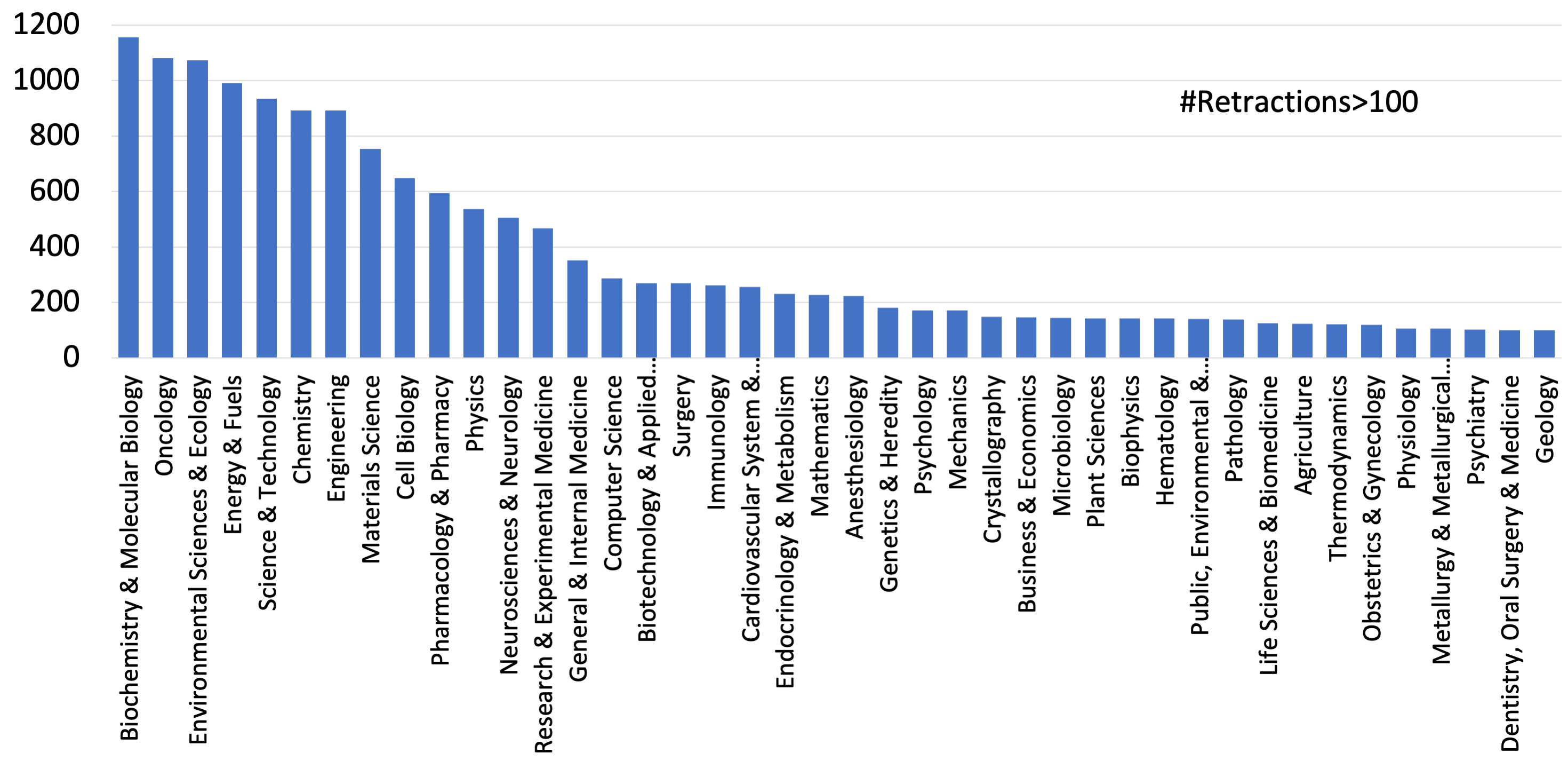}
    \llap{\parbox[b]{5.7in}{(a)\\\rule{0ex}{2.5in}}}
    \includegraphics[width=0.85\linewidth]{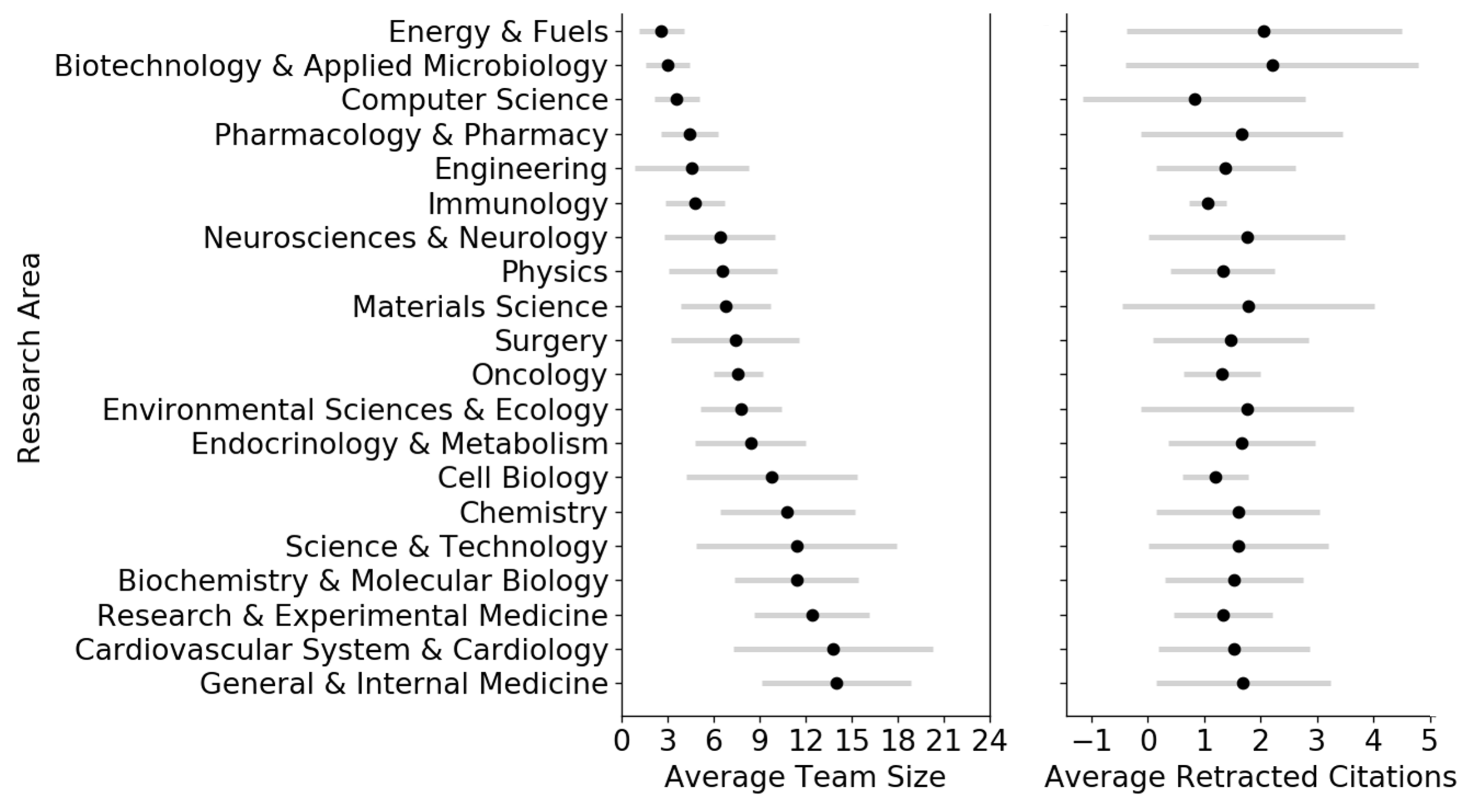}
    \llap{\parbox[b]{3.1in}{(b)\\\rule{0ex}{3in}}}
     \llap{\parbox[b]{1.5in}{(c)\\\rule{0ex}{3in}}}
\caption{(a) Research areas with more than 100 retractions. Results for the first 20 research areas: (b) Average team size (arranged in ascending order); (c) Average number of retracted citations.}
\label{fig:Fig5}    
\end{figure}

Fig.\ref{fig:Fig5}(a) shows research-wise number of retractions (>100).  A large number of retractions are experienced in the biology and oncology discipline. The other categories that have experienced large retractions are ecology and energy disciplines.
The average number of collaborations (team size) for the respective research area is shown in Fig.\ref{fig:Fig5}(b). The biological and medical disciplines have on average large team size.  Fig.\ref{fig:Fig5}(c) shows the average number of retractions respective to the research area. The results show that the journals should perform extra checks on the papers from the disciplines which have experienced a large number of retractions. The disciplines that have a large number of retractions have on average large team size.

\section{Conclusion}
\label{sec:conclusion}
This study provides a number of important insights towards the retractions over four decades.  In addition to a larger sample size encompassing all retractions, this study differs from some previous analyses in terms of the analysis of team size growth, impact of collaborations on retractions, and retracted citations. Teams that are smaller in size have a large number of retractions.
We further have noticed that a handful of authors pair are more active in scientific misconduct and account for a large number of retractions. Also, with the growth of scientific collaborations, the number of retractions is growing. We have analyzed the citations of retracted papers and the very first time have shown the effect of citing the retracted papers on the papers that are citing those retractions. We have found that 1/4th of the retracted papers are such which have received citations and out of those citations, there is at least one paper that turned out to be a retracted paper. This shows that if a fraud is not recognized and retracted on time then how it would affect the new research. Further, to check whether a high-impact retracted paper is highly cited or not, we investigated the relationship between the journal impact factor and the number of retractions, citations on retractions, and retracted citations.  We found that the journal impact factor is independent of the number of retractions and citations. Nowadays, with the increase in the new journals, many low impact journals have lots of retracted papers. A large number of retractions is in a journal with either no impact or low impact.  Our finding showed the effect of citing the retracted paper and that citation has no relation with the journal impact factor. Furthermore, our findings suggest a need for increased attention to the papers that are referring to retracted papers. The rise in the number of retractions raises concern in the scientific community about the degree of responsibility of coauthors and reviewers of scientific articles.
Our results showed that the scientific miscount is independent of the journal impact factor; however, it highlights the journals irrespective of the impact factor that experienced a large number of retractions even after the peer- review.  Both high, as well as low impact journals, experienced retractions that further raise the question of the policies and peer-reviewed process of high impact journals.
It raises an open question to social scientists or other scientific communities to investigate the rationale or psychology behind such frauds.

\section*{Acknowledgment}
We are thankful to Parul Khurana for his help in data extraction and Prof. A.Chakraborti for feedback.

\bibliographystyle{cas-model2-names}

\bibliography{refsBib}

\end{document}